\documentclass[prb,twocolumn,showpacs]{revtex4}
\usepackage{graphicx,epsfig,color}

\begin{document}

\title{Pt-incorporated anatase TiO$_2$(001) surface for solar cell applications :
First-principles density functional theory calculations}

\author{E.~Mete$\,^{a,}$\footnote{Corresponding
author: \indent e-mail: emete@balikesir.edu.tr, \\
secondary address:
Institute of Theoretical and Applied Physics (ITAP)
Turun\c{c}, Mu\u{g}la, Turkey
}, D.~Uner$\,^b$, O.
G\"{u}lseren$\,^c$, and \c{S}.~Ellialt{\i}o\u{g}lu$\,^d$}

\affiliation{$^a$Department of Physics, Bal{\i}kesir University,
\c{C}a\u{g}{\i}\c{s} Campus, Bal{\i}kesir 10145, Turkey \\
$^b$Department of Chemical Engineering, Middle East Technical University,
Ankara, 06531, Turkey \\
$^c$Department of Physics, Bilkent University, Ankara
06800, Turkey \\
$^d$Department of Physics, Middle East Technical University,
Ankara 06531, Turkey}

\date{\today}

\begin{abstract}
First-principles density functional theory calculations were carried
out to determine the low energy geometries of anatase TiO$_2$(001)
with Pt implants in the sublayers as substitutional and interstitial
impurities as well as on the surface in the form of adsorbates. We
investigated the effect of such a systematic Pt incorporation in the
electronic structure of this surface for isolated and interacting
impurities with an emphasis on the reduction in the band gap to
visible region. Comprehensive calculations, for 1$\times$1 surface,
showed that Pt ions at interstitial cavities result in local
segregation, forming metallic wires inside, while substitution for
bulk Ti and adsorption drives four strongly dispersed impurity states
from valence-bands up in the gap with a narrowing of $\sim$1.5 eV.
Hence, such a contiguous Pt incorporation drives anatase into infrared
regime. Pt substitution for the surface Ti, on the other hand,
metallizes the surface. Systematic trends for 2$\times$2 surface
revealed that Pt can be encapsulated inside to form stable structures
as a result of strong Pt--O interactions as well as the adsorptional
and substitutional cases. Dilute impurities considered for 2$\times$2
surface models exhibit flat-like defect states driven from the valence
bands narrowing the energy gap suitable to obtain visible light
responsive titania.
\end{abstract}


\maketitle

\section{Introduction}

The wide-gap semiconductor titania (TiO$_2$) has raised great
interest primarily because of its catalytically active surfaces,
long-standing stability, non-toxicity, and availability of
single-crystal samples.\cite{ollis1,ollis2,fujishima} The most
common phases of titania are known to be anatase, rutile, and
brookite, among which anatase phase proves to be the most promising
for photoelectrochemistry,\cite{chen} visible-light
photocatalysis,\cite{oregan,gratzel,kim} rocking-chair lithium
batteries,\cite{huang} and optoelectronics.\cite{tang1}

Hengerer \textit{et al.}~\cite{hengerer} studied the stability of
anatase TiO$_2$ (101) and (001) facets and found that it is possible
to obtain clean and structurally perfect anatase surfaces. Single
crystals of anatase TiO$_2$ exhibit stronger photocatalytic activity
than rutile phase titania.\cite{thomas,kavan} The natural
crystallographic (001) surface of anatase is most often considered
for catalytic applications among its various facets.\cite{bouzoubaa}

Titania is active under UV irradiation while it is almost inert to
solar spectrum by absorbing only 2\%-3 \% of the sun light. Hence,
narrowing the band gap to visible range is particularly important
for practical photocatalysis. Such applications utilize excess
electrons incorporated by various impurities. Resultant defect
states fall in the band gap of TiO$_2$ sensitizing it for visible
light-induced catalytic purposes. Usual way of defect formation is
achieved by oxygen vacancy which reduces the oxide. However, merely
the vacancy driven defect states are not enough for high level of
activity.

Many attempts have been made to functionalize the titania surfaces
for solar cell applications by impurities in the form of ion doping
or dye sensitization.\cite{kim,kowalska,ko,kitano,wang,nishijima}
Moreover, by acting as charge trap sites, such impurities are
proposed to help in retarding the fast charge-hole recombination
rates, which inherently exist in most of the semiconductors as
TiO$_2$.

Doping TiO$_2$ with metallic as well as nonmetallic elements has
been extensively studied for powdered photocatalysts. Anatase phase
of TiO$_2$ doped with N, S, C, and B has been reported to exhibit
relatively high level of visible
activity.\cite{nishijima,asahi1,sato,umebayashi,irie,zhao}
Supportingly, Wang \textit{et al.},\cite{wang} in their theoretical
study, showed that N doping narrows the band gap of TiO$_2$ by bringing
impurity states in the vicinity of valence-band maximum (VBM). Such
an enhancement can also be obtained by transition-metal ion doping
which reduces the gap allowing visible-light absorption by providing
intraband states near the conduction-band or valence-band
edges.\cite{kowalska,zang,kisch} Yet, it depends on the role of
dopants as recombination centers or as charge traps. For instance,
Co$^{3+}$ and Al$^{3+}$ impurities serve as electron-hole
recombination sites, significantly decreasing the photoreactivity.
Pt ion doping has been successfully shown to enhance visible
activity functioning as charge generation centers which produce free
and trapped charges.\cite{kim,kowalska}

Metal ion doping in the form of Ti substitution has been proposed
also for titania based dye-sensitized solar cells (DSSC) which gain
visible light activation through dye molecule surfactants. In this
manner, DSSC photovoltaic efficiencies have been found to be
remarkably better for doped TiO$_2$ by preventing injected dye
electron recombination between the electrolyte and the
substrate.\cite{ko} On the other hand, Pt incorporation is not
merely limited to powdered photocatalytic systems. Recently, Kitano
\textit{et al.}~\cite{kitano} successfully developed visible-light
responsive Pt-loaded TiO$_2$ thin-film photocatalysts which achieve
separate H$_2$ and O$_2$ evolution from water without requiring dye
sensitization.

Encapsulation of Pt in titania due to the strong metal support
interaction (SMSI) under reducing-gas atmosphere has been reported
by Pesty \textit{et al.}.\cite{pesty} Later, Zhang \textit{et
al.}~\cite{zhang} showed that neutral Pt atoms can thermally diffuse
into TiO$_2$ lattice under oxidizing atmosphere. They also argue
that these diffused Pt atoms can substitute for Ti$^{4+}$ when
oxidized to Pt$^{2+}$ or else they form interstitial impurities as
well.

In this paper, we have investigated the effect of Pt incorporation
in both the lattice and the electronic structure of stoichiometric
anatase TiO$_2$(001) in the form of strongly interacting and
noninteracting impurities on and in the surface. The aim is to
shift the activity to visible region for solar cell operation.
This is done so by band gap narrowing which is useful for
photovoltaic devices. We have considered Pt as adsorbates on the
surface, and as substitutional and interstitial impurities in the
subsurface layers encapsulated by the slab.

\section{Method}

We performed total-energy and electronic-structure calculations
using the Vienna Ab Initio Simulation Package (VASP)
implementation~\cite{vasp} of the gradient-corrected
[Perdew-Burke-Ernzerhof (PBE)] (Ref.~\cite{pbe}) density functional
theory (DFT). The electron-ion interaction has been described by
the projector augmented waves (PAW) method~\cite{paw1,paw2} using
plane-wave basis sets.

The naturally occurring rutile and anatase polymorphs of
titania are basically formed as a result of different modifications
of the same TiO$_6$ unit. This building block is arranged as a
distorted octahedron with a Ti cation at the center and six oxygen
at the vertices. The stacking pattern of these octahedra results in
simple tetragonal ({\it st}) conventional unit cells for both of
rutile and anatase polymorphs. However, in the case of anatase, the
primitive unit cell is a body-centered tetragonal ({\it bct})
Bravais lattice. Since the conventional cell for anatase contains
two {\it bct} units, the calculations assuming an {\it st} unit cell
might cause misleading deductions such as the energy-band-gap type.
We considered bulk properties of anatase phase of TiO$_2$ as the
starting point, both for building up the surface slab models and
for obtaining the bulk-projected electronic structures.
In this manner, we calculated, for example, the bulk lattice
parameters to be $a=3.801$~{\AA}, $c=9.468$~{\AA}, and $u=0.2095$
with a $D_{4h}^{19}$ ($I4_1/amd$) space-group symmetry using {\it
bct} unit cell. These results agree well with the corresponding
experimental values~\cite{howard} which were reported as
$a=3.785$~{\AA}, $c=9.514$~{\AA}, and $u=0.208$. Besides, our
PAW-GGA calculations for these structural properties are consistent
with the other available theoretical
results~\cite{thulin,calatayud1,fahmi,asahi2} and exhibit a better
agreement with the experimental findings.

For the stoichiometric anatase TiO$_2$(001) surface we considered an
oxygen terminated supercell model involving six TiO$_2$ layers with
a vacuum region of $\sim$13 {\AA}. These zigzag-like TiO$_2$ layers
consist of three atomic layers in which bridging oxygen atoms are
out of the level Ti plane. Each of the oxygen at the back surface of
such a bulk termination needs $1/3$ electrons to be saturated. This
cannot be accomplished by hydrogenation with integral charge. Indeed,
it leads hydrogen driven surface states to appear just above the VBM
as a consequence of the excess charge induced by this hydrogenation.
Therefore, instead of saturating the back surface we chose a
virtually symmetrical slab model which produces the same electronic
properties coming from the top and the bottom (001) surfaces. We call
it as virtually symmetrical because it lacks perfect mirror symmetry
along the axis perpendicular to the surfaces. In this sense it is not
a trivial surface to model. On the other hand, our choice bears no
mistakes since we use plane-wave basis sets with periodic boundary
conditions.

In order to elucidate the role of Pt incorporation on the electronic
behavior of anatase TiO$_2$(001) surface, we considered strongly
interacting and isolated impurities on and inside the corresponding
supercells. Pt adsorption on (1$\times$1) surface, for instance,
corresponds to 1 monolayer (ML) coverage so that the shortest
Pt-Pt distance is attained. When one considers Pt with (2$\times$2)
construction, namely, 0.25 ML, Pt-Pt distance becomes 7.53 {\AA} on
the surface which leads to almost isolated impurities.

Because of the symmetrical nature of the slab model, supercell
thickness becomes important particularly when Pt penetration depth
increases. Implanted Pt atoms from the top and back surfaces
should not interact inside the slab. In addition, the central part
of the model slab must possess bulk-like properties. In this manner,
even though none of the atoms were fixed to the bulk positions in
the geometry optimizations, the atoms at the central part retained
their original bulk positions for shallow enough Pt penetration.
Moreover, the number of layers have been chosen so that increasing
the slab height by one more layer did not alter the calculated results
significantly. However, our tests showed that six TiO$_2$ layers are
not enough for Pt impurities placed deeper than 3.4 {\AA} from the
surface oxygen which corresponds to the second TiO$_2$ sublayer.
Therefore, we have used eight TiO$_2$ layer thick slab model for such
Pt impurities inside the slab.

Our convergence tests showed that the electronic wave functions can
be expanded into plane waves up to an energy cutoff value of 400 eV
and that the surface Brillouin-zone integrations can be carried out
with a $k$-point sampling of (8$\times$8$\times$1) and
(4$\times$4$\times$1) Monkhorst-Pack meshes~\cite{mp} for (1$\times$1)
and (2$\times$2) surface unit cells, respectively. In both cases,
these choices gave a total-energy convergence up  to a tolerance value
smaller than 0.1 meV. All geometry optimization calculations were
carried out using conjugate-gradient algorithm based on the reduction
in the Hellman-Feynman forces on each constituent atom to less than
10 meV/{\AA}.

\section{Results and Discussion}

Anatase phase of bulk titania is found to have an indirect gap of
2.08 eV with the valence-band top being at about two-thirds of the
way along $\overline{\Gamma X}$. Direct gap is slightly larger, 2.12
eV. The calculated gap values are underestimated as expected due to
inherent shortcomings associated with the local density approximation
(LDA), while the experimental gap is 3.20 eV.\cite{kowalczyk,tang2}
A more recent work reports this value to be 3.5 eV.\cite{paganini}
The calculational underestimation can be overcome by incorporating
self-energy corrections going beyond the ground-state DFT. Indeed,
Thulin and Guerra~\cite{thulin} reported a quasi-particle
corrected gap of 3.79 eV which is overestimated and closer to the
result 3.68 eV obtained by Calatayud \textit{et al.}~\cite{calatayud1}
using nonlocal B3LYP functional. A comparison of our bulk band structure
(not shown) with the ones reported in these references revealed a
perfect match except the gap width, hence, indicating a scissors
type rigid shift of the whole conduction-band. Therefore,
despite the underestimated gap, the band structures presented in
this work can be considered within this context.

\begin{figure}[ht] \vspace{1mm}
\epsfig{file=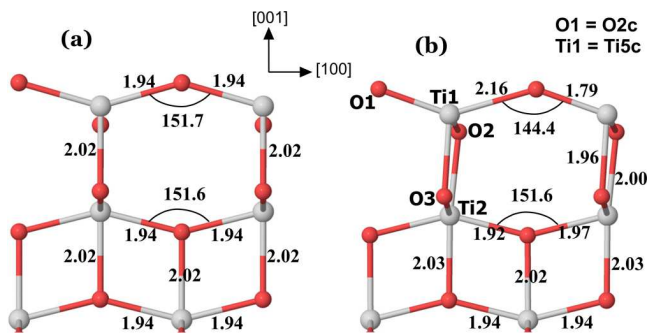,width=8.5cm} \caption{(Color on-line) Atomic
arrangements of the anatase TiO$_2$(001)-2$\times$2 structure. Side
views of the (a) ideal and (b) relaxed surface unit cells, only up to
eight atomic layers are shown here. Ti and O atoms are denoted by
light gray and red (dark) balls, respectively. All bond lengths are
given in angstroms and all angles are presented in degrees.\label{fig1}}
\end{figure}

The (001) plane of anatase polymorph is known to be catalytically
important as it constitutes commercial
catalysts.\cite{martra,devriendt} Yet, a theoretical investigation
on the electronic structure of this surface is still needed.
Therefore, we first consider the oxygen terminated (001) clean
surface. The (1$\times$1) unit cell forms in the shape of a square
with a side of 3.765 {\AA} over this plane [see the top view in
Fig.~\ref{fig3}(a)]. In fact, the (2$\times$2) periodicity has
also similar formation [see the top view in Fig.~\ref{fig5}(a)].
Because of the bulk termination, the surface layer is composed of
fivefold coordinated Ti and twofold coordinated O atoms which are
denoted as Ti5c and O2c in Fig.~\ref{fig1}, respectively. Bridging
O2c atoms make two equidistant bonds with Ti5c's along [100]
possessing a mirror plane symmetry on the ideal surface
[Fig.~\ref{fig1}(a)]. The outward position of O2c forms a
Ti5c-O2c-Ti5c angle of 151.7$^\circ$. In the ideal geometry,
equatorial bond lengths are 1.94 {\AA} while the axial bonds measure
2.02 {\AA}. This conspicuously symmetrical structure gets distorted
upon relaxation in consistency with the previous theoretical
studies.\cite{calatayud2,lazzeri}. The Ti5c-O2c bond lengths become
significantly inequivalent being 2.16 and 1.79 {\AA} mainly because
of the relatively larger displacement of surface oxygen along [100]
compared to the Ti1 and Ti2. Moreover, Ti5c-O2c-Ti5c angle reduces
to 144.4$^\circ$ since O2c's move outward the surface plane while the
relaxation of Ti5c's in the reverse direction is substantially larger
than the other atoms near the surface. This relaxation pattern applies
to both of the surface unit cells as all calculated values related to
the atomic rearrangements presented in Table~\ref{table1} for
(1$\times$1) surface are consistent with those of (2$\times$2). Also,
in agreement with the results of previous theoretical
studies,\cite{calatayud2,lazzeri} our geometry optimizations gave
negligibly small displacements along [010]. Lazzeri \textit{et
al.},\cite{lazzeri} additionally, reported that the planar
O2-Ti1-O3-Ti2 ring becomes slightly skewed making a dihedral angle
of 6.0$^\circ$ with the (100) plane. However, as can readily be seen
in Fig.~\ref{fig1}(b), our calculations suggest that the Ti1-O3 bond
shortens to 1.96 {\AA} and makes an angle of 2.46$^\circ$ with (100)
plane whereas O2-Ti2 bond has a skew angle of 6.0$^\circ$ and a
length of 2.00 {\AA}. This can be attributed to the high reactivity
of the surface with an indication that not only O2c but also
threefold coordinated O2 and O3 give contribution to the surface
electronic properties.

\begin{table}[htb]
\caption{Geometric structure of anatase TiO$_2$(001) surface. Atomic
labels refer to Fig.~\ref{fig1}. Calculated values for atomic
displacements with respect to ideal bulk positions are given in
{\AA}. \label{table1}}
\begin{ruledtabular}
\begin{tabular}{ccrrr}
Surface & Label & [100] & [010] & [001] \\ \hline
1$\times$1 & O1 (O2c) & 0.194 & $-$0.011 & 0.034 \\
& O2 & 0.223 & $-$0.010 & $-$0.038 \\
& O3 & $-$0.117 & $-$0.011 & $-$0.031 \\
& Ti1 (T5c) & 0.003 & $-$0.009 & $-$0.091 \\
& Ti2 & 0.011 & $-$0.011 & $-$0.011 \\ \hline
2$\times$2 & O1 (O2c) & 0.195 & $-$0.011 & 0.034 \\
& O2 & 0.224 & $-$0.010 & $-$0.037 \\
& O3 & $-$0.117 & $-$0.011 & $-$0.030 \\
& Ti1 (Ti5c) & 0.004 & $-$0.010 & $-$0.088 \\
& Ti2 & 0.015 & $-$0.013 & $-$0.012 \\
\end{tabular}
\end{ruledtabular}
\end{table}

An ideal-like structure in which atomic relaxations are only
observed in the direction perpendicular to the surface is found to
be energetically higher by 0.09 and 0.37 eV per unit cell for
(1$\times$1) and (2$\times$2), respectively. The existence of this
symmetry preserving structure was also reported by Calatayud
and Minot~\cite{calatayud2} for (1$\times$1) as well as for
larger unit cells. Minimum energy structure cannot be reached
unless the optimization is started from a slightly distorted ideal
geometry. Relatively small atomic displacements as presented in
Table~\ref{table1} lead to weak energy differences between the
minimum energy structure and the ideal geometry. Therefore, these
weak energy differences can be seen as an indication of high
thermodynamic stability for anatase (001). Indeed we calculated
the surface energies in both of the (1$\times$1) and (2$\times$2)
cases to be 0.92 J/m$^2$ and 0.94 J/m$^2$ for relaxed and unrelaxed
surfaces, respectively. The surface energy is defined as

\[E_{\rm surf} = \frac{1}{2A} \Big(E_{\rm TiO_2}-n \, E_{\rm TiO_2}^{\rm bulk}\Big) \]

\noindent where $E_{\rm TiO_2}$ is the total energy of the slab and
$n E_{\rm TiO_2}^{\rm bulk}$ refers to the energy of the bulk
supercell containing an equal number of TiO$_2$ units as the slab.
$A$ corresponds to the exposed unit-cell area while the factor of
$1/2$ appears because the slab is symmetrical having two faces.
Our calculated values are in good agreement with the GGA-PBE
results ($E^{\rm rel}_{\rm surf}$=0.98 and
$E^{\rm unrel}_{\rm surf}$=1.12 J/m$^2$) of Lazzeri \textit{et
al.}~\cite{lazzeri} and, with the GGA-PW91 results ($E^{\rm
rel}_{\rm surf}$=0.89 and $E^{\rm unrel}_{\rm surf}$=1.04 J/m$^2$)
of Calatayud and Minot\cite{calatayud2}. LDA results are
reported to give systematically larger surface
energies.\cite{lazzeri,hua,oliver} The lack of accurate
experimental measurements to compare with prevents us to discuss
which exchange-correlation scheme does more reliably describe the
surface properties.

\begin{figure}[htb] \vspace{0mm}
\epsfig{file=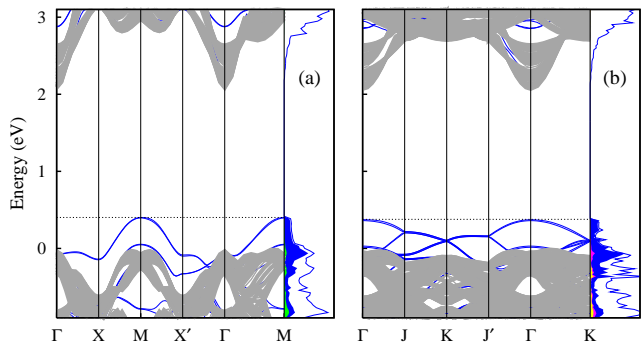,width=8.5cm} \caption{(Color on-line)
Energy bands for the clean anatase TiO$_2$(001) for (a) 1$\times$1
and (b) 2$\times$2 surfaces. Some of the important LDOS
contributions to TDOS are shown. Projection of bulk continuum is
also depicted as shaded areas. \label{fig2}}
\end{figure}

The major surface bands derived from the valence-bands are spilled
out into the energy-band gap. Figure~\ref{fig2} shows the energy bands
for the clean surface of anatase TiO$_2$(001), having (1$\times$1)
and (2$\times$2) periodicities, and the bulk band continuum (shaded
regions) projected on the corresponding surface Brillouin zones,
respectively. Fermi level is at 0.40 eV relative to the valence-band
top and the surface states are filled causing the gap to narrow down
to 1.68 eV (from the bulk gap value of 2.08 eV). These results for
(1$\times$1) surface are in consistency with those of (2$\times$2)
as presented in Fig.~\ref{fig2}(b) and in Table~\ref{table2}. The gap
is indirect with surface state having a maximum at $M$ point for
(1$\times$1) while it is direct with the corresponding maximum at
$\Gamma$ for (2$\times$2) due to the surface Brillouin zone folding.
Conduction band makes a minimum at $\Gamma$ in both cases and is
bulk-like with all the surface solutions being resonance states
within the bulk band region except for the pocket-states. The
right-most panel shows the total density of states (TDOS) and
important local contributions [local density of states (LDOS)] to it.
The surface states in the gap derived mainly from the twofold
coordinated surface oxygen (O2c) are seen to have the form of a
jump discontinuity at $E_{F}$, typical of two-dimensional van
Hove singularity in the DOS due to the critical point at $M$ (or $K$).
The contribution of O2 to the LDOS, being comparably much less than
that of O2c, comes from the lower part of the surface states closer
to the valence-band top. The DOS from O3 (and similarly from Ti5c)
extends to the valence bulk continuum affecting the surface states
even less than O2 does. Evidently, the DOS analysis at the upper
region of the valence-bands indicates that the energy levels of O2c
lying higher than those of fully coordinated oxygen must show higher
reactivity.

We have also calculated the (1$\times$1) and (2$\times$2) clean
surface band structures for the ideal-like geometries in which
Ti5c-O2c bond symmetry is preserved. In this case, the surface
states move upward in the gap region reducing the energy-band gap
by 0.29 eV. Moreover, O2c driven states elevate from the bulk
valence-band up into the gap increasing the number of available
surface states. These elevations are not always rigid. For instance,
the highest lying surface state exhibits some differences between
the ideal-like and the minimum-energy configurations. At $\Gamma$
point, this surface state coincides with the valence-band top in the
lowest energy structure as shown in Fig.~\ref{fig2}(a), whereas it
lies 0.29 eV above the valence-band in the case of the ideal-like
configuration, attesting a prominent change in the character of this
state. Therefore, these differences altogether suggest that the
relaxations of atoms near the surface, noticeably, influence the
surface electronic band structure, in contrary to what was asserted
by a previous theoretical work.\cite{hua}

In practice, titania surface is covered with less than a monolayer
or even dilute metal adatom concentration in catalysis applications.
Hence, we systematically studied Pt implantation in anatase
TiO$_2$(001) with (1$\times$1) and extensionally with (2$\times$2)
periodicities starting from the surface layer up to seven atomic
sublayers corresponding to a depth of $\sim$5.1 {\AA}. First, for
Pt doped TiO$_2$ surfaces, we substitute Pt atoms for the fivefold
Ti atoms (Ti1), closest to the surface layer, and in place of the
bulk Ti atom (Ti2) at the second TiO$_2$ layer. We refer to these
substitutional cases as (s1) and (s2), respectively. Second, we
considered Pt atoms positioned at interstitial sites in oxygen
atomic layers where Pt is strongly coordinated with the
nearest-neighbor oxygen. The trend from strongly interacting Pt
impurities at 1 ML to isolated ones at 0.25 ML coverages is expected
to help predict the more dilute experimental situations.

\subsection{Contiguous impurities}

\begin{figure}[htb]
\epsfig{file=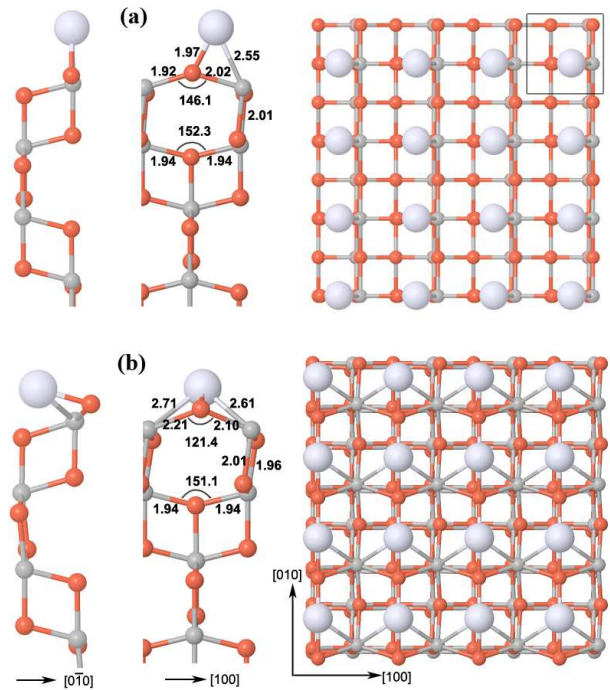,width=8cm} \caption{(Color on-line) Pt on
anatase TiO$_2$(001)-1$\times$1 surface. Front, side, and top views
for the two adsorption cases: (a) on the bridging oxygen bond and (b)
off the bridging oxygen bond. The surface unit cell is indicated in
(a) right. \label{fig3}}
\end{figure}

Pt-implanted TiO$_2$(001)-1$\times$1 system has been considered with
all possible adsorptional, substitutional, and interstitial
configurations starting from above the surface to inside the slab.
Pt impurities in and on the 1$\times$1 surface are separated from
each other by 3.76 {\AA}. Although this is much larger than the Pt
dimer length, it still maintains a distance close enough for a strong
interaction. Pt is found to be stable at two adsorption cases which
are presented in Fig.~\ref{fig3}. In the first case, Pt adsorbate
binds to both Ti5c and O2c at bridge position forming a Ti5c-Pt-O2c
angle of 51$^\circ$. Pt-Ti5c bond length is 2.55 {\AA} which is
significantly close to Pt dimer length while Pt-O2c distance is
1.97 {\AA}. Structural properties of Pt for all the cases studied
in this work such as Pt-O and Pt-Ti bond lengths as well as the
Pt penetration depths are presented in Table~\ref{table2}.
The twofold binding of Pt to Ti5c and O2c compensates their
undercoordination by charge transfer. Hence, the system undergoes
a relaxation in favor of reducing the difference between the two
inequivalent bond lengths (Ti5c-O2c) on the surface layer. (See
Fig.~\ref{fig3}(a)). Similarly, both of the Ti1-O3 and Ti2-O2
bonds get equal in length to a value of 2.01 {\AA} making a
dihedral angle, with the (100) plane, of 5.9$^\circ$ and of
4.6$^\circ$, respectively. Ti5c-O2c-Ti5c angle, on the other hand,
increases slightly to 146.1$^\circ$ due mainly to the downward
relaxation of O2c relative to its nearest-neighbor Ti5c atoms. In
fact, the surface does not show a significant reconstruction upon Pt
adsorption, for this case, and is characterized dominantly by the
relaxation of surface oxygen.

In the second case as shown in [Fig.~\ref{fig3}(b)] Pt is fourfold
coordinated with the surface atoms. It makes two equidistant bonds
with O2c's along [010] being 1.96 {\AA} in length while Pt-Ti bond
distances are 2.71 and 2.61 {\AA}. This small difference in bond
lengths stems from the stronger coordination of Pt with neighboring
O2c's. This strong interaction forms O-Pt-O lines along [010]
as shown in the top view of Fig.~\ref{fig3}(b). In this geometry, Pt
is 0.22 {\AA} closer to the surface relative to the previous case.
Ti5c-O2c bonds are slightly inequivalent being 2.21 and 2.10 {\AA}
similar to the case of the clean surface. Nevertheless,
Ti5c-O2c-Ti5c angle gets subtly reduced to 121.4$^\circ$
prevailingly as a result of the elevation of O2c in a strong
coordination with the Pt adsorbate. The structure in case (b) is
energetically 0.47 eV more favorable than that of case (a) because
of the increased coordination of Pt with the surface atoms. Pt
binding energy (BE) is calculated to be 2.70 and 2.93 eV/atom for
the cases (a) and (b), respectively.

\begin{figure}[htb]
\epsfig{file=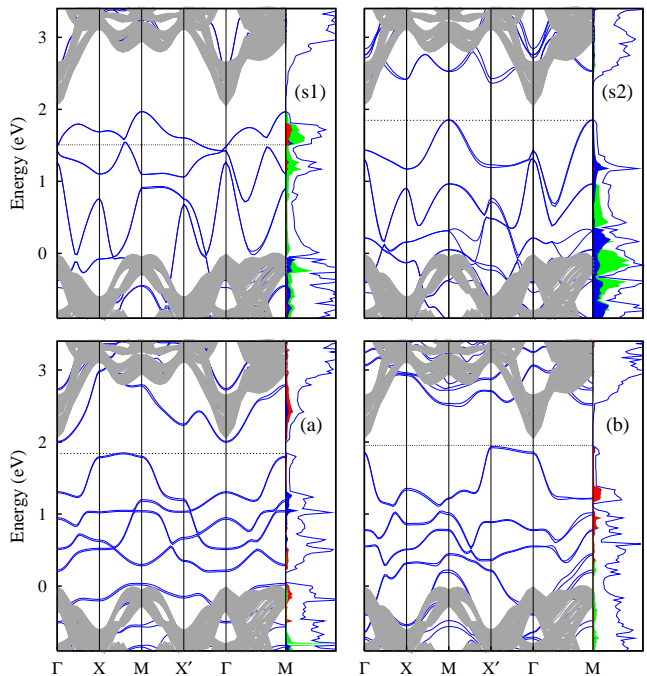,width=8.5cm} \caption{(Color on-line) In
the upper row: (s1) energy bands for Pt impurity atom substituted
for surface-Ti ion in the anatase TiO$_2$(001)-1$\times$1 surface
and (s2) those for Pt impurity atom substituted for the
second-layer Ti ion. Some of the important LDOS contributions to
TDOS are shown. Projection of bulk continuum is also depicted as
shaded areas. In the lower row: (a) energy bands due to
interstitial Pt on anatase TiO$_2$(001)-1$\times$1 surface as in
Fig.~\ref{fig3}(a) and, (b) that as in Fig.~\ref{fig3}(b).
\label{fig4}}
\end{figure}

When Pt is substituted for Ti5c, which is referred as s1, Pt-O3 and
Ti2-O2 axial bond lengths increase substantially to 2.18 and 2.09
{\AA}, respectively. Besides, Pt-O2c equatorial bonds become almost
equal being 1.95 and 1.93 {\AA} with Pt-O2c-Pt angle which reads
152.5$^\circ$. In this structure, Pt-O3-Ti2-O2 side ring as a
whole is skewed making a 5.2$^\circ$ dihedral angle with (100)
plane. Pt is perfectly aligned with nearest-neighbor oxygen in the
same lanes over the surface along [100] and [001] directions while
being at the different atomic layers.

On the other hand, Pt substitution for bulk Ti (Ti2), namely,
s2, results in a structure which reflects similar topological
characteristics with the ideal clean slab [shown in Fig.~\ref{fig1}(a)].
Pt-Ti2 replacement at 1$\times$1 unit cell drives the structure
from the relaxed to an ideal-like geometry. Even though Pt replaces
Ti2, the axial bond lengths extend slightly to 2.02 {\AA}. Similarly,
the difference in bond distances of Pt with the neighboring fully
coordinated oxygen become negligibly small being 1.95 {\AA}.
The only substantial displacement with respect to the relaxed
structure is obtained for O2c which moves up a little bit so that
Ti5c-O2c bond lengths become symmetrized with a value of 1.96 {\AA}
forming an isosceles  Ti5c-O2c-Ti5c triangle having an obtuse angle
of 146.9$^\circ$.

Pt can also be considered at the interstitial sites in between the
fully coordinated level oxygen. Starting from such a configuration,
the surface expansively reconstructs with a local segregation as a
result not only of the SMSI, but also of the stress induced by the web
of closely spaced Pt impurities inside the slab. For instance, when
we place Pt in between two O2 atoms (referring to Fig.~\ref{fig1}),
it pushes the first TiO$_2$ layer (O1-Ti1-O2 group) upward
breaking Ti1-O3 and Ti2-O2 bonds. Meanwhile, Pt moves in between
O1 and O3 forming new equidistant axial bonds (1.98 {\AA}) on a
straight line which makes an angle of 26.6$^\circ$ with the (100)
plane. Since O1 and O2 coordination numbers interchange, O2 is
further elevated up forming zigzag Ti5c-O2c pattern, this time
along [010] direction with bond lengths of 1.85 and 2.08 {\AA}. In
summary, Pt interstitial at O2 level for (1$\times$1) surface
segregates the first and the second TiO$_2$ layers by 2.0 {\AA}
relative to the separation of those in the clean surface.

The coordination of Pt with nearest-neighbor oxygen in these model
cases signifies the strength of Pt-O electrostatic interaction so
that Pt is very dominant in disturbing the surface atomic
arrangements. These impurity driven rearrangements modify the
electronic band structure to a significant extent by bringing new
defect states as well as perturbing the already existing surface
states that originate from the distorted lattice bondings.

The electronic structure for the geometries in Figs. 3(a) and 3(b)
with Pt ions as surface impurities are presented in Figs. 4(a) and
4(b), respectively. Moreover, in Fig. 4(s1) and 4(s2), those for the
substitutional impurity at the surface (replacing Ti1) and that in
the bulk (replacing Ti2 in the subsurface layer) are shown,
respectively.

%Fig. 4(a):
For the first case of adsorbates shown in Fig. 4(a), the energy gap
is full of six defect states, five of which are occupied and one that
is close to the conduction-band is empty. Fermi level is at 1.84 eV.
The system has a narrow gap of only 0.15 eV. The gap is indirect and
the highest occupied defect state makes a maximum at midway,
otherwise almost flat-band, along $\overline{XM}$. The lowest
unoccupied band is also a defect state making a minimum at $\Gamma$
and having the shape of the lowest conduction-band along most
directions except around $X'$. Its LDOS has a two-dimensional
character as expected, and is a result of antibonding interactions
between electrons from Pt and O1 ions. The lower group of four
bands all have Pt contributions with three upper bands having
contributions from O1. Especially, the flat-band at 1.0 eV causing a
peak in the LDOS and the peak above that are mainly due to O1 ion,
whereas the lowest of the four has contributions mainly from O2 ion.
This band and the one below, separated by a pseudo-gap, which is barely
above the valence-band top, are of the same character, namely, of Pt
and O2.

%Fig. 4(b):
The second adsorbate case described in Fig. 3(b) is energetically
more preferable than the first one as they both contain equal
number of atomic species. In this case [see Fig. 4(b)] the same
five defect states fill the bulk band gap as in the previous case.
All of them lie below the Fermi level which is at 1.95 eV relative to
valence-band top and the gap is 0.13 eV, again small. The lowest
unoccupied state being bulk-like this time is the conduction-band
minimum at $\Gamma$. The empty defect state is higher in energy. The
highest occupied defect state has a maximum at $X'$, however the
band is almost flat along $\overline{X'\Gamma}$ making the gap
nearly a direct one. The band is also flat in some part of the way
along $\overline{\Gamma M}$, both causing higher densities in
the LDOS picture. The contribution by the O1 ion is mainly in the
upper defect bands in the energy range between $E_F$ and about 0.2
eV below that along with Pt ion, while that by O2 is mainly in the
lower defect states starting from the highest peak at about 0.6 eV
down into the valence-bands.

%Fig. 4(s1):
In the case of surface substitutional, the bulk energy gap is full
of four defect states as seen in Fig. 4(s1), lower two of which are
fully occupied and the upper two are almost half-filled each.
Therefore the system shows metallic behavior. Fermi level is at
1.51 eV relative to the valence-band top. These two upper bands
cross each other along high-symmetry directions of the surface
Brillouin zone at three places, namely, very close to the
$\Gamma$ point along $\overline{X'\Gamma}$, and passed midway
along the $\overline{\Gamma M}$ and $\overline{XM}$ directions.
The LDOS due to Pt ion as well as some due to neighboring two
oxygen ions O1 and O2 are coming from these bands dispersed around
the Fermi level. The two lower bands are surface states due to
oxygen ions O1 and O2.

%Fig. 4(s2):
Figure 4(s2) shows the band structure for the subsurface substitutional
where the energy gap is full of four defect states that are all
occupied. Fermi level is at 1.85 eV relative to the valence-band top
giving rise to an indirect energy gap of 0.23 eV. The LDOS of the
highest occupied defect band having a sharp peak due to the almost
flat part along $\overline{X'\Gamma}$ is contributed by Pt ion
and its neighboring oxygen ions. This band is pretty similar in
shape to the half-filled one in (s1) case that makes a maximum at $M$.
Next piece of LDOS between 0.3 and 0.9 eV is due to the surface
oxygen O1. This band was observed to be similar to the clean
surface band due to O1 in Fig.~\ref{fig2}(a), especially around $M$
point along all three directions. The band that is hardly visible
above the valence-band top at $M$ point is also similar to the second
band of clean surface at the same region. Since the Pt ion is deeper
than skin of the surface region in this case the surface is heeled
and the surface states are back. The van Hove singularity at about
0.9 eV, where this band ends at the right-most $M$ point in
Fig.~\ref{fig4}(b), is also evident just like the one at $E_F$=0.40
eV in Fig.~\ref{fig2}(a).

\begin{table*}
\caption{Calculated values of some key parameters for
Pt-TiO$_2$(001) anatase system: work function,
Fermi energy relative to bulk valence top, and change in energy-band
gap, as well as Pt-depth relative to surface oxygen, Pt-O distance,
and Pt-Ti distance for each model.
Labeling of models follow from Figs~\ref{fig3} and \ref{fig5}, for
(1$\times$1) and (2$\times$2) surfaces, respectively.
\label{table2}}
\begin{ruledtabular}
\begin{tabular}{cc|ccc|ccc}
Surface & Model & $W$(eV) & $E_{F}$(eV) &
$\Delta E_{\rm g}$(eV) & $h_{\rm Pt}$(\AA) & $d_{\rm Pt-O}$(\AA) & $d_{\rm
Pt-Ti}$(\AA) \\ \hline
1$\times$1 & Clean & 6.88 & 0.40 & ~~\,0.00 & ~~\,--- & --- & --- \\ %E_g=1.68
& (s1) & 5.83 & 1.51 & ~~\,--- & $-0.46$ & 1.94(O2c),1.96(O2),2.18(O3) & --- \\
& (s2) & 5.36 & 1.85 & $-1.45$ & $-3.04$ & 1.95(O3,O4),2.02(O2,O5) & --- \\ %E_g=0.23
& (a)  & 4.16 & 1.84 & $-1.53$ & ~~\,0.76 & 1.97(O1) & 2.55(Ti1) \\ %E_g=0.15
& (b)  & 5.21 & 1.95 & $-1.55$ & ~~\,0.54 & 1.96(O1) & 2.61(Ti1), 2.71(Ti1) \\[-4mm] \\ \hline %E_g=0.13
2$\times$2 & Clean & 6.77 & 0.38 & ~~\,0.00 & ~~\,--- & --- & --- \\ %E_g=1.70
& (s1) & 6.13 & 1.12 & $-1.13$ & $-0.68$ & 1.92(O2c), 2.10(O2c), 1.99(O3) & --- \\ %E_g=0.57
& (s2) & 6.25 & 0.88 & $-0.50$ & $-3.04$ & 1.99 (O3,O4), 2.02 (O2,O5) & --- \\ %E_g=1.20
& (a)  & 5.86 & 0.96 & $-0.58$ & ~~\,1.24 & 1.90(O1) & 2.36(Ti1) \\ %E_g=1.12
& (b)  & 5.61 & 1.41 & $-1.03$ & ~~\,0.24 & 1.95(O1),2.07(O3) & 2.30(Ti1),2.78(Ti1) \\ %E_g=0.67
& (c)  & 5.64 & 1.80 & $-1.42$ & ~~\,0.02 & 1.99(O1) & 2.80(Ti1) \\ %E_g=0.28
& (d)  & 5.71 & 1.51 & $-1.13$ & $-2.91$ & 1,97(O3),2.06(O3) & 2.72(Ti1),2.63(Ti2),2.86(Ti2) \\ %E_g=0.57
& (e)  & 5.69 & 1.49 & $-1.11$ & $-3.59$ & 2.01(O4) & 2.71(Ti2),2.77(Ti2),2.91(Ti3) \\ %E_g=0.59
& (f)  & 5.62 & 1.56 & $-1.18$ & $-5.29$ & 2.01(O5) & 2.88(Ti2),2.75(Ti3) %E_g=0.52
\end{tabular}
\end{ruledtabular}
\end{table*}

\subsection{Dilute impurities}

\begin{figure*}
\epsfig{file=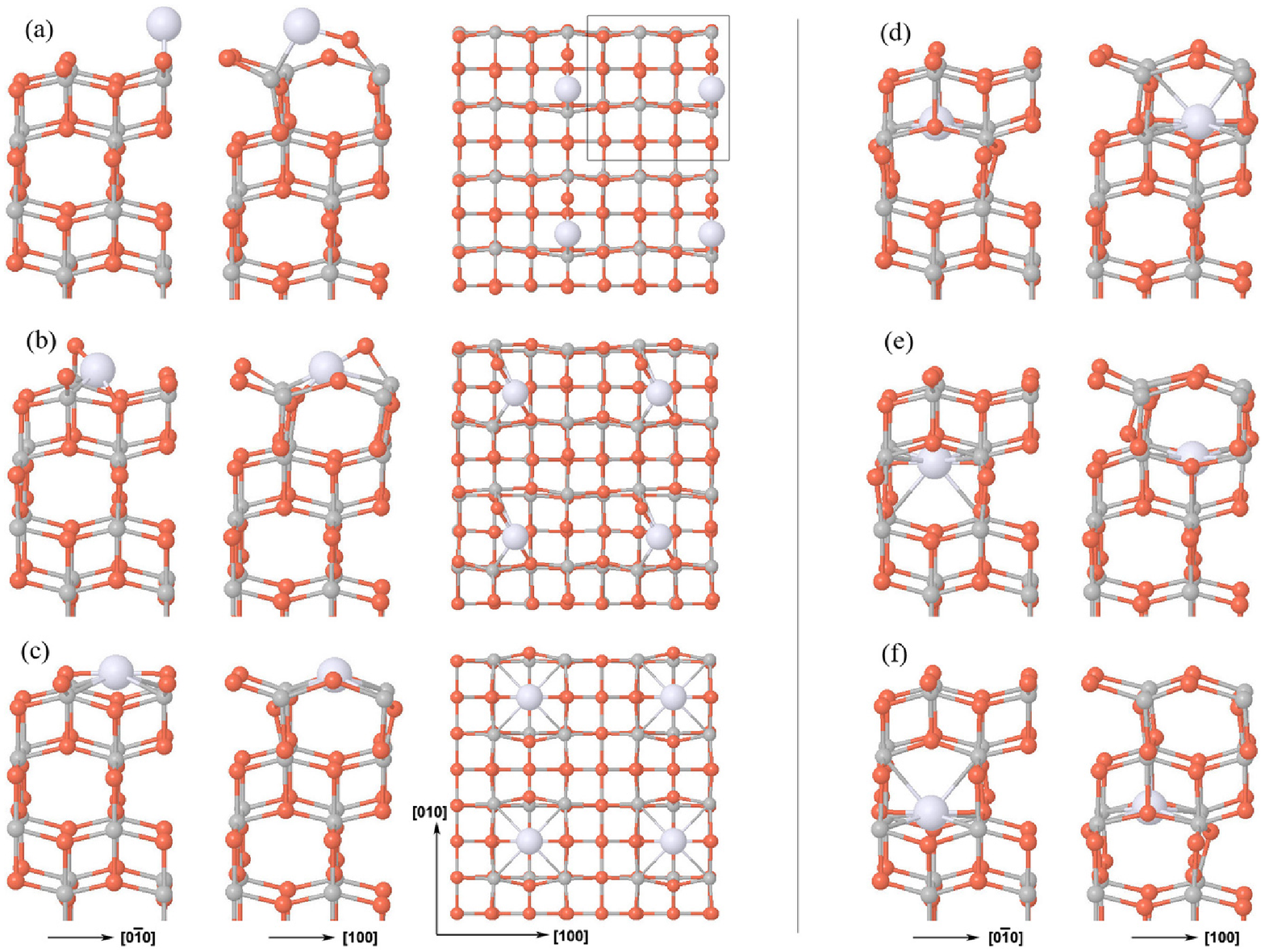,width=17cm} \caption{(Color on-line)
Interactions of single Pt atom with anatase TiO$_2$(001) surface (in
order of penetration depth of the Pt atom): (a) Pt adsorbed on the
surface, (b) surface oxygen supported by Pt with SMSI, and [(c)-(f)]
stable structures of Pt in the surface. Pt atom is shown in white
and denoted by the biggest sphere while Ti is in gray and O is
represented as small red (dark) balls. \label{fig5}}
\end{figure*}

TiO$_2$(001)-(2$\times$2) unit cell prevents charge transfers
between the impurity sites by providing a 7.53 {\AA} Pt-Pt
separation together with local screening effects through the slab.
Therefore, both the geometric and the electronic structures show
significant differences from the (1$\times$1) counterparts. For
instance, Pt interstitials can be stable inside (2$\times$2) slab
without segregation effects as opposed to the case with (1$\times$1)
surface. For the incorporation of Pt ion as an adsorbate or as
an interstitial, we examined a total of six cases that are shown in
Figs.~\ref{fig5}(a)-\ref{fig5}(f), in the first three of which the Pt
is situated on/in the surface and in the last three the Pt ion is below
the surface layer.

In the first case, [Fig.~\ref{fig5}(a)], Pt adsorbate is twofold
coordinated with nearest-neighbor Ti5c and O2c along the O1-Ti1
row. Pt raises the O2c upward by 0.92 {\AA} and pushes it in [100]
direction by 0.83 {\AA} from its relaxed lattice position. As a
result of the SMSI, bond distances from Ti5c to Pt, Pt to O2c, and
O2c to the next Ti5c, which are connected successively through line
segments in a row, become 2.36, 1.90, and 180 {\AA}, respectively.
The other O1-Ti1 row remains to be less effected by the presence of
Pt adsorbate with Ti5c-O2c bonds, 2.11 and 1.81 {\AA} in length,
making a Ti5c-O2c-Ti5c angle of 144.4$^\circ$ as in the case of
relaxed clean surface. Among the side bonds, other than the skewing
toward [$\bar 1$00], merely the Ti1-O3 one, which connects to the
promoted O2c, is elongated by 0.04 {\AA} with a dihedral angle of
7.23$^\circ$ bearing a subtle difference from the values calculated
for the (2$\times$2) clean surface. Moreover, the distances and
angles are not distorted considerably in the second TiO$_2$ layer
near the surface.

The structure shown in Fig.~\ref{fig5}(b), being the lowest energy
configuration among adsorptional and interstitial cases, undergoes a
more complex atomic rearrangement upon Pt adsorption. The adsorbate
migrates to a bridge position in between Ti1 and O2 with bond
lengths of 2.78 and 2.07 {\AA} for Pt-Ti1 and Pt-O2, respectively.
As a consequence of the SMSI, the nearest-neighbor O2c is pushed
from its lattice site upward above the midpoint between Ti1 and Pt
forming a triangle. It is aligned perpendicular to the surface, with
O2c-Ti1 and O2c-Pt sides of 1.81 and 1.95 {\AA}, respectively. In
this geometry O2c elevation above the Pt adsorbate is calculated to
be 0.83 {\AA}. Another triangle forms between the consecutive Ti1 on
the Ti1-O1 row, neighboring O2 and Pt which lies at the vertex
connecting the two triangular atomic arrangements in a fourfold
coordination with these surface atoms. In this second triangle
Pt-Ti1 bond distance becomes 2.30 {\AA}. Furthermore, the two
Ti5c-O2c bonds become slightly distorted, 1.81 and 2.17 {\AA} in
length, with Ti5c-O2c-Ti5c angle of 143.9$^\circ$ on the second
Ti1-O1 row. The only noticeable difference occurs in the Ti1--O3
bond which extends to 2.16 {\AA}. All other side bonds preserve
the skewing similar to the (2$\times$2) clean surface results.

The third adsorptional model adopts a more neatly symmetrical
structure among Pt/TiO$_2$(001)-(2$\times$2) cases as shown in
Fig.~\ref{fig5}(c). Pt resides at the midpoint between two
undercoordinated surface oxygen with a bond length of 1.99 {\AA}.
Meanwhile, Pt also makes equidistant bonds with four Ti5c's, each of
which 2.8 {\AA} in length. This extended bond distance (see
Table~\ref{table2}) entails a rather weak Pt-Ti5c interaction on
the surface. As a result of this isotropic fourfold coordination
with surface Ti's, the axial Ti1-O3 bonds become aligned parallel
to [001] direction. On the other hand, the two Ti2-O2 bonds, being
coplanar with the Pt, are the only slanted ones which make two
dihedral congruent angles of 12.3$^\circ$ with the (100) and
($\bar{1}$00) planes.

When Pt is substituted for one of the surface Ti ion (Ti1), the
Pt-O3 bond, although unaltered in length, noticeably differs from
Ti-O3 bonds by skewing more from 2.21$^\circ$ to 18.21$^\circ$. Due
to the excess charge incorporated by the substitutional impurity,
the Pt-O2-Ti1 row exhibits slight distortions in bonds and angles.
Resultant two Pt-O2c-Ti5c angles become 141.8/153.5$^\circ$ with
Pt-O2c and Ti5c-O2c bond distances of 1.92/2.10 and 2.01/1.82
{\AA}, respectively. On the other hand, the Ti1-O2-Ti1 row
retains the values as those of the clean surface.

As in the case of (1$\times$1), Pt substitution for Ti2 results in
the reduction in the skewness in the axial bonds as they align
parallel to [001] direction. While the axial bond lengths assume
an ideal-like value of 2.02 {\AA}, the nearest-neighbor equatorial
Pt-O distances become 1.99 {\AA} which describes a slightly
extended value over the Ti-O equatorial bond length of 1.94 {\AA}.
Therefore, the only substantial displacement is obtained  for the
atoms at the surface TiO$_2$ layer along [$\bar{1}00$]. This
rearrangement reduces the skewness but preserves the non-symmetric
nature of Ti5c-O2c bonds which read 2.14 {\AA} and 1.81 {\AA} with a
Ti5c-O2c-Ti5c angle of 144.8$^\circ$. In other words, as the number
of Pt substitutions for four possible Ti2 ions increases, the structure
adopts an ideal-like geometry which gains a plane mirror symmetry as in
the case of s2-$1\times$1.

For anatase TiO$_2$(001)-(2$\times$2) surface, being contrary to the
(1$\times$1) cases, Pt is found to be stable once it penetrates into
the interstitial cavities, starting from the 03 level, as shown in
Figs.~\ref{fig5}(d)-\ref{fig5}(f). Pt ions are encapsulated by the slab
in an octahedral position at the midpoint between the two level oxygen
in interaction with the six nearest-neighbor Ti's. The location of the
encapsulated Pt is determined dominantly by the strong Pt-O
interaction, as being at the same depth with and in the middle of,
the fully coordinated two level oxygen. This leads to two structural
ramifications as fingerprints. First, two oxygen atoms whose
interconnecting line is perpendicular to O-Pt-O bonding, in the
closest (preceding or succeeding) oxygen atomic layer, are slightly
repelled out from their lattice positions as a result of the induced
stress due to the excess charge brought by the impurity. This is not
solely specific to interstitial cases. We obtained the very similar
result for the adsorptional model in Fig.~\ref{fig5}(c). Second, Pt
ion, making equidistant equatorial bonds with, maintains four Ti's
at the corners of a square shape which can be seen through [001]
direction. This is due mainly to Pt impurity rather than the TiO$_2$
lattice itself since the third adsorptional case [see top view of
Fig.~\ref{fig5}(c)] adopts surface Ti's in the same geometry. This is
so that Ti1-O3 axial bonds are aligned parallel to [001] as opposed
to their slanted posture in the clean surface. Evidently, similar
arguments apply for the first two interstitial cases
[Figs.~\ref{fig5}(d) and~\ref{fig5}(e)], as well. This time Pt holds
Ti2's in a squarely manner causing Ti2-O2 bonds to align vertically
with respect to the surface plane. When Pt penetrates deeper than the
second TiO$_2$ layer into the cavities, the symmetry breaking over the
Ti5c-O2c bonds and the skewness of Ti1-O3-Ti2-O2 ring, similar to the
clean surface can finally be reproduced. This corresponds to the
last interstitial model as shown in Fig.~\ref{fig5}(f). Therefore, Pt
implants induce a local stress causing nearby atoms to slightly
rearrange from their lattice positions in the TiO$_2$(001)-(2$\times$2)
surface as a result of the strong Pt-O interaction.

\begin{figure}[htb]
\epsfig{file=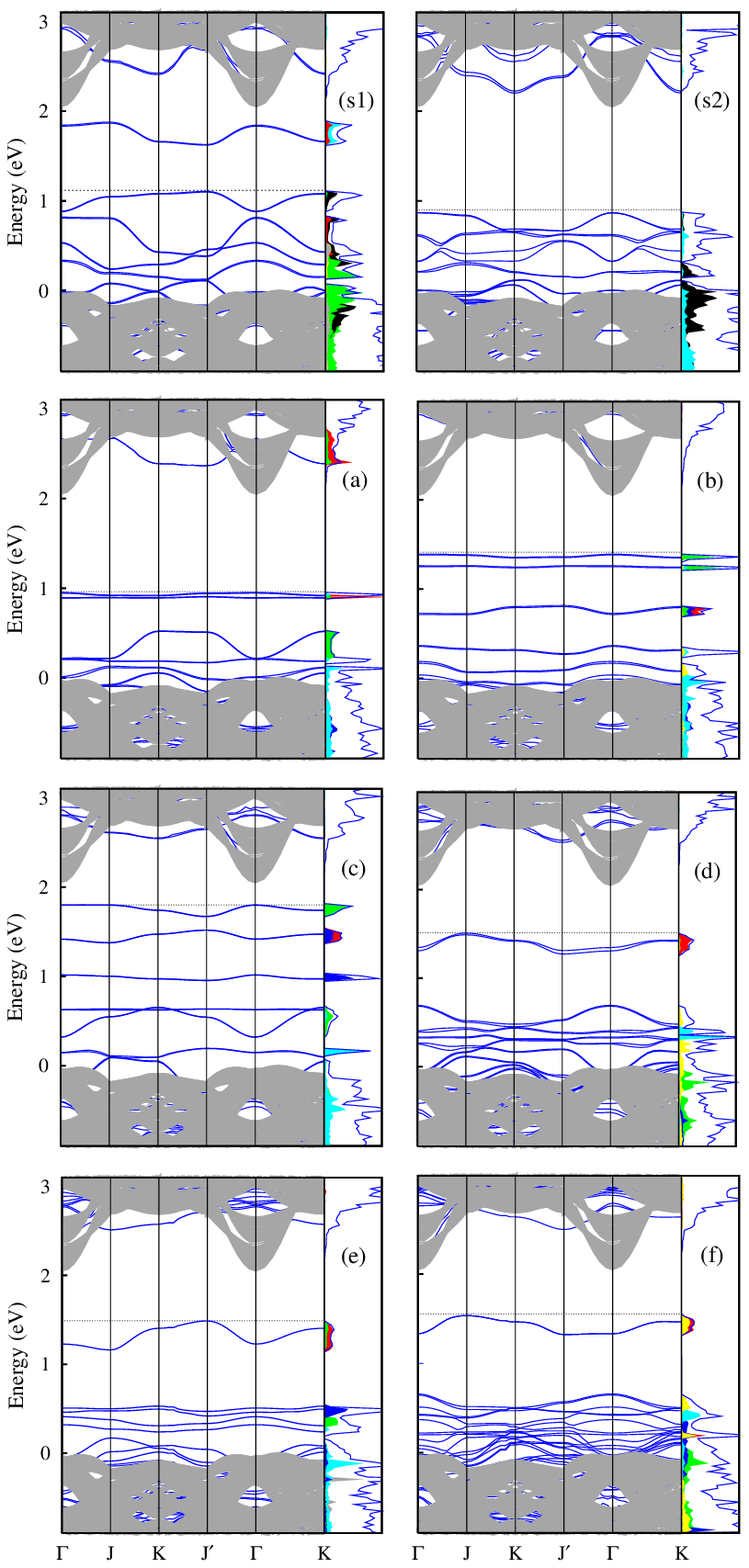,width=8.5cm} \caption{(Color on-line)
(s1) Energy bands for the anatase TiO$_2$(001)-2$\times$2 surface
with Pt adsorbed as a substitutional impurity at a surface-Ti
site, (s2) at a subsurface-Ti site, and [(a)-(f)] for
the interstitial impurity cases in Figs.~\ref{fig5}(a)-\ref{fig5}(f).
\label{fig6}}
\end{figure}

The electronic structure for the geometries in
Figs.~\ref{fig5}(a)-\ref{fig5}(f) with the interstitial impurity are
presented in  Figs.~\ref{fig6}(a)-\ref{fig6}(f), respectively, and
moreover in Figs.~\ref{fig6}(s1) and~\ref{fig6}(s2), those for the
substitutional impurity at the surface (replacing Ti1) and that in
the bulk (replacing Ti2 in the subsurface layer) are shown,
respectively.

%Fig. 6(a):
For the first adsorbate case described in Fig.~\ref{fig5}(a), the
impurity bands due to the interaction between Pt and the neighboring
promoted-oxygen O1$'$ (prime for bonded neighbor) are grouped in three
sets within the bulk band-gap region [see Fig.~\ref{fig6}(a)]. Two of
them being almost conjugate of each other are in the energy ranges of
2.3--2.6 and 0.15--0.5 eV, respectively, while the bands in the
third group, being almost flat, producing a sharp peak in the LDOS
picture, are located in between the other two groups at around 1 eV.
This is the highest occupied set of states of mainly Pt-O1$'$
character with some contributions from the O2s. The Fermi level is at
0.96 eV, and since the lowest unoccupied band is the bulk
conduction-band having a minimum at $\Gamma$ the energy gap is direct
and 1.12 eV in width. The conjugate bands are dispersed along $k_y$ but
flat along $\Gamma J$ and $KJ'$. The empty one is in resonance with the
bulk bands around $\Gamma J$ and localized in the gap otherwise. Due to
this asymmetry in localization the LDOS is also not symmetric in shape.
The filled one being in the gap has a symmetrical one-dimensional (1D)
LDOS  coming from Pt-O1$'$ interaction, with a smaller mixture of Pt
and  larger mixture of O1$'$, which is the other way around in the
empty conjugate state. In addition to bonding to O1$'$, Pt causes
slightrepositioning to other surface oxygen, as a result of which, a
little contribution comes from O1 along the same line as Pt-O1$'$
[see Fig. 4(a)] in the upper part of the delocalized 1D-peak. The
lower peak of the same LDOS is degenerate with a contribution due to
a new flat band of O2 character adjacent to the edge of the above
band. The lower two bands are surface-like and again of O2
characters.

%Fig. 6(b):
The case shown in Fig.~\ref{fig5}(b) is the most favorable one with
the lowest total energy among all (except the substitutional cases
since they belong to different stoichiometry which prevents a direct
comparison of their total energies) and its electronic nature is
again a semiconductor, like all the other dilute impurity cases.
Comparing its energy bands [see Fig.~\ref{fig6}(b)] with the case of
Fig.~\ref{fig6}(a), the empty impurity band (not shown) is now pushed
up into the bulk continuum of the conduction-band. The two flat peaks
(Pt-O1$'$), split by 0.15 eV, are also raised in energy causing the
Fermi level to be at 1.41 eV, and decreasing the (direct) energy gap
to 0.67 eV. They have also contributions from O1 and O2$'$. The next
lower band (due to Pt-O1$'$-Ti1-O2 chain) is now much less dispersed
and it is separated from the impurity band below, being no longer
degenerate at $\Gamma$ as in case of Fig.~\ref{fig6}(a). The fifth
(shown) band is also due to Pt-O1$'$ and Ti1--O1 interactions.

Figure~\ref{fig5}(c) shows the last of the adsorbate cases where Pt
is situated very symmetrically on the O1 layer. All four Ti1's are
equivalent; two of the four O1's are bonded to the Pt interstitial,
and two of the four O2's are displaced as seen in top view of
Fig.~\ref{fig5}(c). Consequently, there are eight bands fallen into
the gap. The first one from the top, again empty, is within the
conduction-band, with localization around $K$, and conjugate to the
sixth band. It is mainly of Pt character. The second band is the
highest occupied O1$'$-related surface band which makes a maximum at
$\Gamma$ and flat along $\Gamma J$. The Fermi level is at 1.80 eV and
the (direct) gap is 0.28 eV, the narrowest gap of all cases. The third
band at about 1.5 eV is an impurity band due to Pt-O2$'$ interaction of
larger Pt mixing, whereas the fourth band, at about 1 eV, has a sharp
peak of same character with larger O2$'$ mixing.
The fifth band is a perfect flat one due to displaced O2s since
Pt pushes O2s off their lattice positions causing these stress-induced
flat going states that reflect major oxygen character due to relatively
weaker neighboring interactions.
The sixth band is a rather dispersed two-dimensional surface band of
O1$'$, O2, and O1 characters, in decreasing order of contribution,
which is almost the mirror image of the empty first band above. They
are both very symmetric with respect to $k_x$ and $k_y$ as expected
[see top view of Fig.~\ref{fig5}(c)]. The last two, the seventh and
eighth bands are impurity bands of mainly O2$'$, O2, and slightly of
O1$'$ and Pt contributions.

%Fig. 6(d):
Since the geometric structures, shown in Figs.~\ref{fig5}(d)-\ref{fig5}(f),
of the Pt ions as bulk interstitials (subsurface and deeper) in their
nearest-neighbor environment are equivalent, their electronic band
structures are also very similar, especially for (d) and (f). It is
expected that as the Pt ion is placed deeper into the bulk [see
Figs.~\ref{fig6}(d)-\ref{fig6}(f)], the band structure will alternate
as ``(d) and (e)". The empty defect band lies partly in the gap. The
second impurity band being filled is due to Pt-O1$'$ (O2, O3$'$, O4$'$)
interactions, and making a maximum at $J$, $J'$, and J in cases (d),
(e), and (f), respectively. Similarly, the Fermi level is situated
at about 1.5 eV and the energy gap is about 0.5 eV (see Table II).
The lower filled bands are the mixture of surface bands and less
dispersed impurity bands due to Pt interstitials.

%Fig. 6(s1):
When the surface Ti is substituted by a Pt impurity on a 2$\times$2
reconstruction, the gap is filled by several impurity bands as seen
in Fig.~\ref{fig6}(s1). One of them is far above in the conduction-band
region and partly fallen into the gap. The next lower one is an
empty band making a minimum at $J'$ point around $\sim$1.6 eV. The
bandwidth is about 0.2 eV due to the interaction between the Pt
substitutional impurity with its neighbors O3$'$ and O1$'$. Below is
the highest occupied band of surface-like state dispersed by the
interaction between surface Ti and O2$'$. This band makes a maximum,
$E_F$=1.12 eV, at $J'$ point as well, causing the 0.57 eV gap to be
a direct one and narrowed as compared to the 1.70 eV gap for the
clean surface. The band is nearly flat along $\overline{JK}$ and
$\overline{KJ'}$ which results in a sharp peak in the LDOS at
Fermi level. Having the upper empty band going almost flat along
$\overline{KJ'}$, as well causing parallel bands in this
direction, one may expect an enhancement in the optical transitions
rate. The two bands further below are due to O3$'$-Pt-O1$'$ and
Ti--O2$'$ bonding interactions, respectively. Morover the bands below
those look more like, roughly, the surface states of clean (001).

%Fig. 6(s2):
In the case of placing the substitutional Pt ion at the second layer
we have a bulk-like impurity problem where the surface layer is
chemically similar to the clean surface with four Ti ions exposed.
This is also evident in the band structure shown in Fig.~\ref{fig6}(s2).
The empty states fallen from the conduction-band into the forbidden
gap, but still above the conduction-band minimum, are very symmetric
with respect to $k_x$ and $k_y$ directions. And in the optical gap,
most of the bands look surface-like with impurity bands passing
through them. The Fermi level is at 0.88 eV and the gap of 1.20 eV
is again a direct one at $\Gamma$ like in the clean surface case.
The difference of 0.50 eV is partly (about 0.30 eV) due to the
un-reconstruction of the clean surface with the subsurface impurity
substitution. The Pt based impurity state is almost flat at around
0.2 eV above the valence-band top.

The strong dispersion of defect bands seen in (1$\times$1) cases and
the rather less dispersed nature of these bands in (2$\times$2)
models originating from the Pt impurities signify the role of Pt-Pt
interaction in relation to Pt concentration. Furthermore, the
distinguishable flatness of the impurity bands shown in
Fig.~\ref{fig6}(a) and~\ref{fig6}(b) arise from the minimal
coordination of Pt adsorbate with the surface ions as a result of its
spatial location, in addition to the diluteness of these impurities.

1 ML Pt substitution for undercoordination drive anatase into a
metallic state. At the same coverage Pt adsorption, and substitution
for the second-layer Ti in this surface yields a low band-gap
semiconducting system which would be active in the infrared
region. On the other hand, anatase TiO$_2$(001) can be
functionalized for visible activity by Pt impurities implanted with
a (2$\times$2) periodicity for all models except the (c) case [see
Fig.~\ref{fig5}(c)] which corresponds to a band gap narrowing of
1.42 eV. For this impurity concentration, this represents the
maximum value close to those of the 1 ML cases. LDOS analysis
indicates that the number of defect states derived from the valence
bands increase with the coordination of Pt with O1 and O2 ions which
is also maximum due to its high-symmetry relaxed position. Therefore,
even though, platinized TiO$_2$ is known to give visible-light
activity,\cite{kim} our calculations show that band gap narrowing
depends on the impurity concentration and the coordination number of Pt
with the near surface oxygen.

\subsection{Analysis of the electronic density}

The topological analysis of the charge density gives accurate
information about the bonding characteristics for the neighboring
atoms. Therefore, one needs a qualitative description of the
interatomic charge distributions which can be computed by employing
Bader analysis based on atom in molecule (AIM) theory. To do so,
the real-space cell is partitioned into Bader volumes delimited by
local zero-flux surfaces of the electron density gradient vector
field. Then, these volumes can be integrated around an atomic region
to calculate the local charge depletion and accumulation. We obtained
these atomic properties using the AIM formalism that is implemented
with a recent grid-based algorithm.\cite{sanville} The Bader charge
results for Pt and surface region Ti and O atoms are presented in
Table~\ref{table3} in the cases of (1$\times$1) and (2$\times$2) unit
cells. For (2$\times$2) surface, there are four possible Ti (or O)
ions to choose from which are at the same atomic layer. We have
preferred to provide the calculated values for the ones that are
closer to the Pt impurity, in order to make a better comparison with
the (1$\times$1) counterparts.

\begin{table}
\caption{The valence charge accumulation based on Bader analysis
for Pt-TiO$_2$(001) anatase system. Atom labels follow
Fig.~\ref{fig1}. The charges for lattice atoms that are closest to
Pt are presented in the case of (2$\times$2) slab.
\label{table3}}
\begin{ruledtabular}
\begin{tabular}{cc|cccccc}
Surface & Model & O1 & O2 & O3 & Ti1 & Ti2 & Pt \\ \hline
1$\times$1 & Clean & $-$1.26 & $-$1.34 & $-$1.33& +2.61 & +2.64 & --- \\
& (s1) & $-$0.79 & $-$0.99 & $-$1.15 & --- & +2.62 & +1.63 \\
& (s2) & $-$1.27 & $-$1.20 & $-$1.02 & +2.57 & --- & +1.84 \\
& (a) & $-$1.15 & $-$1.34 & $-$1.32 & +2.52 & +2.64 & -0.03 \\
& (b) & $-$1.17 & $-$1.33 & $-$1.31 & +2.39 & +2.64 & +0.11 \\[-4mm] \\ \hline
2$\times$2 & Clean & $-$1.26 & $-$1.34 & $-$1.33 & +2.60 & +2.64 & --- \\
& (s1) & $-$1.07 & $-$1.07 & $-$1.33 & +2.58 & +2.63 & +1.64 \\
& (s2) & $-$1.26 & $-$1.19 & $-$1.18 & +2.60 & +2.64 & +1.85 \\
& (a) & $-$1.07 & $-$1.35 & $-$1.33 & +2.49 & +2.64 & +0.02 \\
& (b) & $-$1.08 & $-$1.31 & $-$1.33 & +2.57 & +2.65 & +0.06 \\
& (c) & $-$1.22 & $-$1.32 & $-$1.33 & +2.54 & +2.64 & +0.07 \\
& (d) & $-$1.26 & $-$1.33 & $-$1.23 & +2.56 & +2.56 & +0.08 \\
& (e) & $-$1.23 & $-$1.34 & $-$1.27 & +2.58 & +2.55 & +0.15 \\
& (f) & $-$1.26 & $-$1.34 & $-$1.31 & +2.60 & +2.59 & +0.14
\end{tabular}
\end{ruledtabular}
\end{table}
\begin{figure}[ht] \vspace{2mm}
\epsfig{file=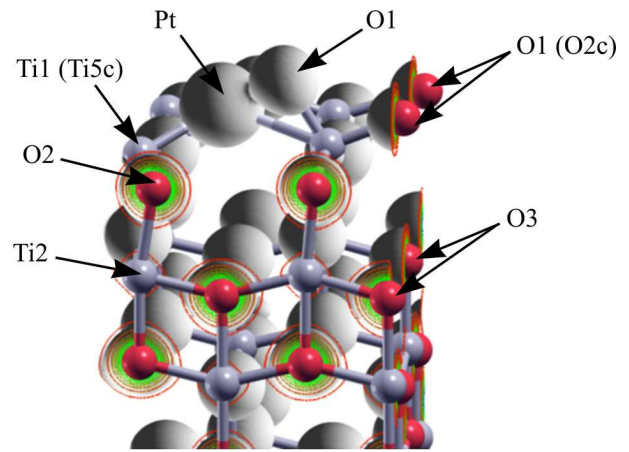,width=8cm} \caption{(Color on-line) 3D
charge-density plot for the ``b-2$\times$2" case of Pt on anatase
TiO$_2$(001) surface [Fig.~\ref{fig5}(b)]. \label{fig7}}
\end{figure}

In the case of clean surface, a deep lying Ti ion, which should
reflect bulk-like properties, transfers an amount of 0.443
electronic charges to each of the neighboring oxygen losing its
last atomic shell. Hence, this fully coordinated Ti ion accumulates
a Bader charge of $+2.66e$. A bulk-like oxygen, on the other hand,
gets a valence charge state of $-1.33e$. Calatayud
\textit{et al.}~\cite{calatayud1} computed these values for the
bulk anatase  TiO$_2$ as $Q_{\rm Ti}=2.96e$ and $Q_{\rm O}=-1.48e$
using a different exchange-correlation scheme. The Bader charge
accumulation around the ions is also sensitive to the determination
of integration regions with boundaries delimited by zero gradient of
the electronic density. Although we obtained slightly lower charge
states for Ti an O ions resulting in a relatively less polarized
bonding, in quite an agreement both results obey the same
stoichiometry by correctly adding up to charge neutrality of TiO$_2$
and are smaller than the nominal oxidation states obtained for a
generic ionic oxide such as MgO. As a result, a polarized covalent
bonding develops between charged Ti cations and O anions.

Naturally, the AIM charge values, presented in Table~\ref{table3},
for (1$\times$1) and (2$\times$2) periodicities in the case of clean
surface models exhibit a one-to-one correspondence as expected, since
the electronic properties (such as the band gap, work function, etc.)
derived from their charge densities must represent the same surface. The
oxidation states of surface layer ions affect the reactivity of single
crystals of anatase (001). Hence, we computed the Bader charges for the
undercoordinated ions as $-1.26e$ for O1 and $+2.61e$ for Ti1 being
lower  by $\sim 5\%$ and $\sim 1\%$ than that of the deep lying O and
Ti, respectively. Moreover, bulk-like charge states are adopted starting
from O2 atomic layer which stays $2.56$~{\AA} below the surface oxygen.
Clearly, bulk termination bears slight differences in charge states
of deep and surface ions, particularly in the case of surface oxygen,
indicating a rather low reactivity of the clean (001) surface. This
prediction is in good agreement with the experimental observations that
clean surfaces of TiO$_2$ exhibit lower catalytic activity than stepped
(101) and oxygen-defect (001) surfaces.\cite{thomas}

Pt incorporation yields significant disturbance in the electronic
density in the vicinity of the impurity site. Particularly, this
effect is observed for the oxygen in the close proximity of Pt
while the electron depletion from around the nearest-neighbor Ti
ion remains minimal upon Pt deposition. From the data presented
in Table~\ref{table3}, the standard deviation in the Bader charges
with respect to the reference clean surface values have been
calculated to be 0.17, 0.14, and 0.12 for O1, O2, and O3 while
the corresponding values are 0.09 and 0.04 for Ti1 and Ti2,
respectively. Smaller deviations obtained for the deeper lying
ions also imply a limited contribution of these atoms to
the DOS and surface bands near the Fermi level. Therefore,
for instance, we obtain the smallest band gap for the adsorptional
case (c) among (2$\times$2) structures although it has lower number
of Pt-O interactions than interstitial configurations do.

When substituted for Ti ion either at the surface or in the slab,
Pt can not acquire the same oxidation state that Ti had. This
results in a relatively lower amount of charge accumulation around
the neighboring oxygen resulting in a less polarized covalency
between Pt and O. This is clear also, for instance, in the
three-dimensional (3D) charge-density plot presented in
Fig.~\ref{fig7}, which belongs to the case (b) of Pt-TiO$_2$
(2$\times$2) system. Ti's show lesser valence electronic density
distributions around them, indicating strongly polarized covalent
Ti-O bonding. The electron depletion from Pt to O1 is lower than,
that, for instance, from Ti2 to neighboring oxygen. Moreover, the
charge density around O1 is noticeably smaller than that of O3 (and
also of O2). In summary, Ti-O bond polarization, and therefore
interaction, in TiO$_2$ lattice environment is stronger than that
of Pt-O.

On the other hand, the strength of Pt-TiO$_2$ interaction can also
be compared with respect to different model cases presented in
Table~\ref{table3}. Bader charges calculated for Pt ion suggest
that it interacts with the lattice in the substitutional cases
stronger than in the adsorptional and interstitial configurations.

\subsection{Thermodynamic stability of the phases}

The (1$\times$1) and (2$\times$2) surface supercells comprise
unequal amounts of atomic species. Moreover, adsorptional or
interstitial cases, being stoichiometrically different from the
substitutional ones, represent an addition of impurity instead of
a Pt-Ti replacement. When the supercell total energies are considered,
the lowest energy structures are b-1$\times$1 [Fig.~\ref{fig3}(b)]
and b-2$\times$2 [Fig.~\ref{fig5}(b)] among the adsorptional/interstitial
cases while they turn out to be s1-1$\times$1 [Fig.~\ref{fig3}(s1)] and
s2-2$\times$2 [Fig.~\ref{fig5}(s2)] for the substitutional geometries
at one and quarter ML concentrations, respectively. Therefore one
cannot directly compare their stability by just looking at their
supercell total energies.

We employed formalism of Qian \textit{et al.}~\cite{qian} and
Northrup~\cite{northrup} to study the thermodynamic stability of the
Pt-incorporated TiO$_2$(001) surfaces that have varying number of
constituents at different concentrations. In this scheme, relative
formation energy is defined as a function of the chemical potential
of the excess atomic species as
\[
E_{\rm form}=E_{\rm Pt/TiO_2}-E_{\rm TiO_2}-\Delta n_{\rm Ti}\mu_{\rm Ti}
-\Delta n_{\rm Pt}\mu_{\rm Pt}
\]
where $E_{\rm Pt/TiO_2}$ and $E_{\rm TiO_2}$ are the total energies
of the Pt incorporated and bare TiO$_2$ surfaces while $\mu_{\rm
Ti}$ and $\mu_{\rm Pt}$ stands for the chemical potentials of Ti and
Pt. $\Delta n_{\rm Ti}$ and $\Delta n_{\rm Pt}$ represent the
differences in the number of atoms of each atomic species with
respect to the reference clean surface. Formation energy in this form
is a function of Ti and Pt chemical potentials. Equilibrium is reached
when the chemical potential of a given species is equal in all the
phases that are in contact with each other. Also, these phases must
be in equilibrium with bulk anatase such that
\[
\mu_{\rm Ti}+2\mu_{\rm O}=\mu_{\rm TiO_2}.
\]

\begin{figure}[tbh] 
\epsfig{file=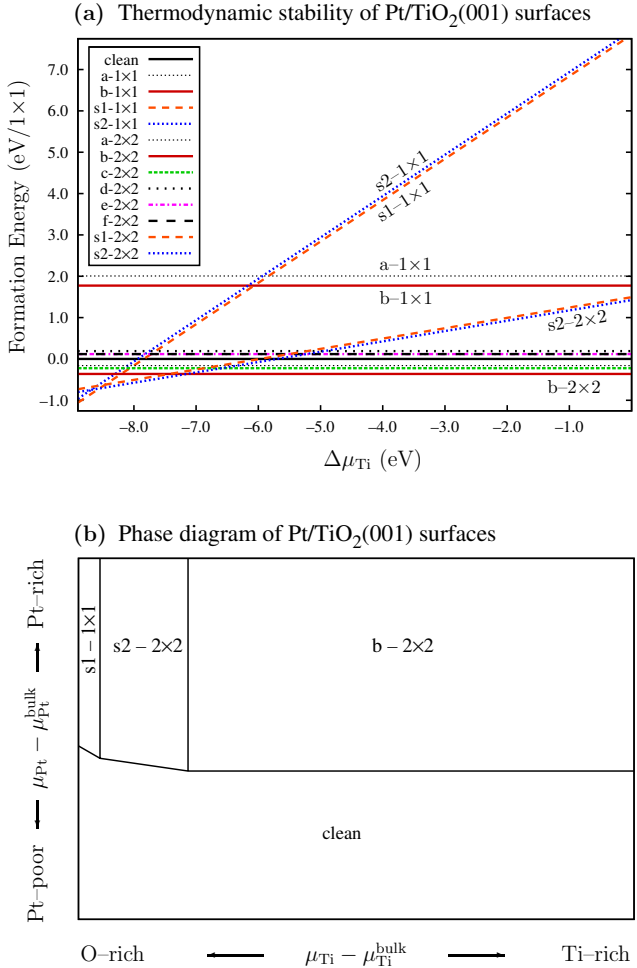,width=8.4cm} \caption{(Color on-line)
(a) Normalized formation energies of Pt--incorporated TiO$_2$(001)
structures relative to that of clean surface as a function of the Ti
chemical potential. (For numerical figures, see Table~\ref{table4})
Pt impurities are chosen to be in thermodynamic equilibrium with
$fcc$ Pt bulk phase. (b) Phase diagram of Pt/TiO$_2$(001) surface
as a function of Pt and Ti chemical potentials.
\label{fig8}}
\end{figure}

This relation interrelates chemical potential of Ti to O chemical
potential that varies accordingly with the experimental conditions.
The value of $\mu_{\rm Ti}$ must be smaller than that of the $hcp$
Ti bulk solid phase which is an undesirable formation at the
surface and is referred as the Ti-rich conditions. The other extreme
is that when the surface oxygen are found in thermodynamic equilibrium
with the molecular oxygen bath corresponding to O-rich conditions.
Assuming it as an ideal gas, the intermolecular interactions can be
neglected. Hence, the chemical potential of oxygen can be referenced to
the value at the O$_2$ molecule  which is $\mu_{\rm O}=E_{{\rm O}_2}/2$.
Therefore, Ti chemical potential relative to its bulk value
$\Delta\mu_{\rm Ti}=\mu_{\rm Ti}-\mu_{\rm Ti}^{\rm bulk}$ varies between
$\Delta\mu_{\rm Ti}=0$ (Ti-rich conditions) and $\Delta\mu_{\rm Ti}=-8.9$ eV
(O-rich conditions), for TiO$_2$. Since the relative formation energy
of phases is defined as a function of Ti and Pt chemical potentials,
Pt is assumed to be in thermodynamic equilibrium with its $fcc$ bulk solid
phase that serves as a reservoir for Pt atoms. Hence, $\mu_{\rm Pt}$ is
chosen to represent such an experimental condition. In fact, this
corresponds to just another extreme for undesired Pt phases on the
surface that can be avoided by $\mu_{\rm Pt}<\mu_{\rm Pt}^{\rm bulk}$.

\begin{table}[ht]
\caption{Numerical values for the formation energy of Pt/TiO$_2$
surfaces for adsorptional and interstitial phases (all in
eV/1$\times$1) referring to the constant valued functions depicted
in Fig.~\ref{fig8}. Relative formation energies are given with
respect to that of the bare surface. Those of the substitutional
phases not provided since they vary with varying Ti chemical
potential due to different surface stoichiometry.
\label{table4}}
\begin{ruledtabular}
\begin{tabular}{ccc}
Phase & $E_{\rm form}$ (Absolute) & $E_{\rm form}$ (Relative) \\[1pt] \hline
Clean         & 0.76 & ~~\,0.00 \\
a-1$\times$1 & 2.76 & ~~\,2.00 \\
b-1$\times$1 & 2.53 & ~~\,1.77 \\
a-2$\times$2 & 0.59 & $-0.17$ \\
b-2$\times$2 & 0.39 & $-0.37$ \\
c-2$\times$2 & 0.53 & $-0.23$ \\
d-2$\times$2 & 0.94 & ~~\,0.18 \\
e-2$\times$2 & 0.87 & ~~\,0.11 \\
f-2$\times$2 & 0.87 & ~~\,0.11 \\
\end{tabular}
\end{ruledtabular}
\end{table}

The formation energies are shown in Fig.~\ref{fig8}(a) relative
to that of the clean surface for 12 phases as a function of Ti
chemical potential over the full range of its allowed values.
They are normalized to a 1$\times$1 unit cell in order to compare
their thermodynamic stabilities. Pt adsorption at 1 ML coverage
results in unstable surfaces with relative formation energies at
1.77 and 2.00 eV/1$\times$1 for (b) and (a) adsorption modes,
respectively. Substitutional cases at this concentration are the
most unstable phases within the range of $-8.56<\Delta\mu_{\rm Ti}<0$
eV with increasing instability toward Ti-rich conditions while they
become the most stable phases under O-rich conditions for
$-8.90<\Delta\mu_{\rm Ti}<-8.56$ eV in favor of substitution for Ti5c
(s1-1$\times$1). For this latter experimental situation, in which the
surface is in thermodynamic equilibrium with the molecular oxygen,
s2-1$\times$1 phase has a relatively higher formation energy by
0.10 eV/1$\times$1 followed by s2-2$\times$2 and s1-2$\times$2
structures which are energetically less favorable by 0.26 and 0.33
eV/1$\times$1. Similarly, over the range $-7.15<\Delta\mu_{\rm Ti}<0$ eV,
Pt substituted TiO$_2$(001)-(2$\times$2) surfaces are the most
unstable structures among 0.25 ML phases whereas they turn out
to be the most stable cases, in which the formation energy is
slightly lower for s2-2$\times$2 phase, over the range
$-8.56<\Delta\mu_{\rm Ti}<-7.15$ eV closer to O-rich conditions.

On the other hand, Pt interstitials are unstable relative to the
formation energy of the clean surface by 0.18 eV/1$\times$1 for (d),
and by 0.11 eV/1$\times$1 for (e) and (f) at (2$\times$2)
reconstruction implying increased stability with increasing Pt
penetration depth. These results suggest that practical interstitial
applications might require thermal treatment. Therefore, our
calculations for the stabilities of substitutional cases under
O-rich conditions and of Pt interstitial phases are in good agreement
with the experimental results of Zhang~\textit{et al.}~\cite{zhang}
who suggested that neutral Pt atoms can thermally diffuse into TiO$_2$
lattice under oxidizing atmosphere. They also argue that these
diffused Pt atoms can either substitute for the Ti$^{4+}$ sites
when oxidized to Pt$^{2+}$ (for which our calculations show a charge
state of $+1.85e$) or they occupy interstitial sites.

In contrary to those of the 1 ML phases, Pt adsorbates at 0.25 ML
happens to be energetically more stable relative to bare surface by
0.17, 0.23, and 0.37 eV/1$\times$1 for the (a), (c), and (b) adsorption
phases, respectively, for the whole range of allowed Ti chemical
potential. Besides, b--(2$\times$2) structure is also the most stable
phase within $-7.15<\Delta\mu_{\rm Ti}<0$ eV which spans 80.3\% of
the whole range from Ti-low to Ti-rich conditions.

The phase diagram shown in Fig.~\ref{fig8}(b) has been derived from
the results obtained for the energetically more stable phases,
presented in Fig.~\ref{fig8}(a), for varying Ti and Pt
chemical-potential values. Hence, high formation energy surfaces have
not been considered due to their instability. Under Pt-rich conditions,
which refer to the formation energies presented in Fig.~\ref{fig8}(a),
three most stable phases exist for varying chemical potential of Ti.
Under Pt-rich and O-rich conditions the most stable phase is
s1-1$\times$1 in which all of the surface-Ti ions substituted with
Pt. A small deviation from these conditions by slightly increasing
the Ti chemical potential switches the phase to 0.25 ML concentration
surface of s2-1$\times$1. For lower O concentrations b-2$\times$2
surface is more stable and dominant over the range of allowed Ti
chemical potential. Under Pt-poor conditions the phase diagram
reproduces the clean TiO$_2$(001) surface with no impurities.

\section{Conclusions}

We systematically studied the structural and electronic properties
of Pt impurities in the form of adsorptional, interstitial, and
substitutional cases for anatase TiO$_2$(001) with (1$\times$1)
and (2$\times$2) surface periodicities. The former represents full
coverage while the latter corresponds to isolated impurities.
Depending on the Pt concentration per unit cell area,
impurity-impurity electron coupling strength mediates the mode
of atomic rearrangements as they are clearly different for
(1$\times$1) and (2$\times$2) models. For instance, Pt adsorption
at the bridge site obtained for 1 ML coverage in Fig.~\ref{fig3}(a)
is corresponded by the pattern shown in Fig.~\ref{fig5}(a) at 0.25 ML
coverage, in which surface oxygen is promoted by the adsorbate
as a result of the SMSI. This difference is even more pronounced
for interstitial cases. When implanted inside the slab for full
coverage, Pt atoms form parallel metallic wires inside TiO$_2$
where interlayer distances slightly increase due to local
segregation, while Pt impurities can be encapsulated by the
(2$\times$2) lattice at interstitial cavities to form
structures without undergoing a major reconstruction.

In addition, another dominant factor in the formation of low
energy Pt/TiO$_2$ structures is the nearest-neighbor Pt-O
coordination which derives from the impurity-lattice oxygen
charge transfer. Relative stabilities of these structures can
be addressed to local disturbance on the potential-energy
surface induced by the excess electrons brought by the impurities
that consequently account for the enhancement of the electron
trapping efficiencies.

Clean TiO$_2$(001) possesses surface states derived from the
valence-bands in the energy-band gap near the VBM originating
from the undercoordinated surface oxygen. The nature of these
bands is sensitive to the minimum energy rearrangement of
the surface ions. Although this accounts for the observable
responsiveness of TiO$_2$(001), electronic charge-density
analysis indicates a relatively lower reactivity with respect
to defect surfaces.

In interaction with nearest-neighbor oxygen, Pt derives
$s$-$p$ hybrid impurity states from the valence-bands, which
lie above the VBM close to the Fermi level, as well as the
ones from the conduction-bands. The character of these defect
states controls the narrowing of the energy-band gap. In addition
to the Pt-O interaction, impurity-impurity coupling, depending
on the Pt concentration, influences the position and the
dispersion of defect states. Therefore, 1 ML cases exhibit
strongly dispersed bands in the gap with a narrowing that
refers to infrared activity and ultimately to metallization
for Pt-Ti1 substitution and for interstitial Pt implantation.
Having the highest symmetry adsorption mode, Pt at the midpoint
between two O2c's at 0.25 ML also exhibits a band gap
narrowing of 1.42 eV corresponding to infrared region similar
to low band-gap Pt/TiO$_2$(001)-(1$\times$1) cases. Pt
concentration, therefore, is not the only factor that controls
the band gap narrowing. Our calculations show that the number
of impurity driven states that fall into the gap is proportional
to the coordination of Pt with O1 and O2. This explains the
occurrence of 1.42 eV gap narrowing for 0.25 ML Pt concentration
while all the remaining cases for (2$\times$2) represent visible
activity. Moreover, excess non-bonding charge due to the
interaction of Pt with only one O1 in the adsorptional models
in Figs.~\ref{fig5}(a) and~\ref{fig5}(b) brings rather flat going
bands.

At 1 ML concentration under oxygen rich conditions Pt
substituted TiO$_2$ tends to be more stable. As O chemical
potential gets slightly lower, the phase adopts a (2$\times$2)
surface periodicity. Further decreasing $\mu_{\rm O}$ toward Ti-rich
conditions, the most stable phase, dominant over the allowed
range of experimental conditions, happens to be adsorption on
the surface for 0.25 ML coverage (b-2$\times$2), which is
semiconducting in the visible region with a band gap narrowing
of 1.03 eV. Consequently, anatase TiO$_2$(001) can be
functionalized for visible activity by Pt incorporation. We
have shown how it relates to impurity concentration, to
coordination number of Pt with the near surface oxygen and
to thermodynamic conditions which determine the chemical
potential of constituent species, for useful applications.

\section{Acknowledgements}

E.M. and {\c{S}}.E. acknowledge financial support from T\"{U}B\.{I}TAK,
The Scientific and Technological Research Council of Turkey (Grant No:
TBAG 107T560), and O.G. acknowledges the support of Turkish Academy
of Sciences, T\"{U}BA. We are also grateful to Middle East Technical
University for providing further support through Projects No.
BAP-2004-07-02-00-100 and No. YUUP-BAP 2004-08-11-06.

\end{document}